\documentclass[3p,times,11pt]{article}

\usepackage{graphicx}
\usepackage[utf8]{inputenc}
\usepackage{amssymb}
\usepackage{amsthm}
\usepackage{amsmath}
\usepackage{epstopdf}
\usepackage{amsfonts}
\usepackage{subfigure}
\usepackage{psfrag}
\usepackage{float}

\usepackage{xcolor}
\usepackage{tikz}
\usepackage{pstricks}

\usepackage{fancyhdr}
\usepackage{hyperref}

\usepackage{comment} 

\usepackage{setspace}
\usepackage{cite}
\usepackage{color}
\usepackage[left=2.5cm, right=2.0cm, top=5.0cm, bottom=2.5cm]{geometry}

\title{Relativistic hyperbolic motion and its higher order kinematic quantities}
\author{Iván Pérez-Román, Haret C. Rosu}

\date{\textit{\small{IPICyT, Instituto Potosino de Investigación Científica y Tecnológica,\\
Camino a la presa San Jos\'e 2055, Col. Lomas 4a Secci\'on, 78216 San Luis Potos\'{\i}, S.L.P., Mexico}}\\[\baselineskip] November 2, 2022}

\begin{document}

\maketitle

\begin{abstract}
We investigate the kinematics of the motion of an observer with constant proper acceleration (relativistic hyperbolic motion) in $1+1$ and $1+3$ dimensional Minkowski spacetimes. We provide explicit formulas for all the kinematic quantities up to the fourth proper time derivative (the Snap).
In the $1+3$ case, following a recent work of Pons and de Palol [Gen. Rel. Grav. 51 (2019) 80], a vectorial differential equation for the acceleration is obtained which by considering constant proper acceleration is turned into a nonlinear second order differential equation in terms of derivatives of the radius vector. If, furthermore, the velocity is parameterized in terms of hyperbolic functions, one obtains a differential equation to solve for the argument $f(s)$ of the velocity. Differently from Pons and de Palol, who employed the particular solution, linear in the proper time $s$, we obtain the general solution and use it to work out more general expressions for the kinematical quantities. As a byproduct, we obtain a class of modified Rindler hyperbolic worldlines characterized by supplementary contributions to the components of the kinematical quantities.
\end{abstract}

{\em Keywords:} hyperbolic motion, Jerk, Snap, Minkowski spacetime, modified Rindler hyperbolas\\

Revista Mexicana de F\'{\i}sica  68, 060702 (2022), DOI: https://doi.org/10.31349/RevMexFis.68.060702\\

\section{Introduction}

In this paper, we study the kinematical quantities of accelerated motion in Minkowski spacetime, in which, as well known, one of the coordinates is the laboratory time. In the lowest dimensional (one time component and one spatial coordinate) Minkowski spacetime, the relativistic motion of constant proper acceleration (the hyperbolic motion)  has been throughly discussed over more than a century, starting with Born \cite{born1909theory}.
Even when only considering uniform proper acceleration, intriguing quantum field effects, like the famous Unruh effect, show up \cite{unruh1976notes,crispino2008unruh}.

In the recent literature on this topic, there have been discussions related to particular features of uniform and non-uniform relativistic acceleration \cite{russo2009relativistic}, in which case, higher order time derivatives of the four-velocity, like the Jerk, $J(s)$, Snap, $S(s)$, Crackle, $C(s)$, and beyond, should be considered. 
Acceleration and further higher-order proper time derivatives become important and non-trivial quantities in generalizations related to intrinsic differential geometric parameters of the curves (worldlines) like in \cite{letaw1981stationary}, or related to curved spacetimes as in \cite{friedman2015uniform} and \cite{paithankar2019bound}.\\

The organization of the bulk of this paper is the following. In Section 2, we briefly present the kinematics of the hyperbolic motion in the $1+1$ Minkowski case with an iterative extension to higher order proper time derivatives.
In Section 3, we consider the hyperbolic motion in four-dimensional Minkowski spacetime. In this case, a matrix representation of Lorentz transformations between a traveler proper frame and a laboratory frame can be used, see e.g. \cite{pons2019observers}. In this matrix approach, one obtains a nonlinear second order differential equation which is solved based on the standard parametrization of the four-velocity in terms of hyperbolic functions. The four-acceleration is calculated, together with its modulus. Similar calculations are performed for the Jerk and Snap. In Section 4, the general solution of the differential equation for hyperbolic parametrization is obtained and the same kinematical quantities are expressed in term of the general solution. In Section 5, the worldlines (Rindler-type hyperbolas) based on the general solution are obtained and displayed graphically. Section 6 contains the conclusions of this work.

\section{Hyperbolic motion in the 1+1 dimensional Minkowski spacetime}

The relativistic hyperbolic motion happens when a particle moves in a Minkowski spacetime with constant proper acceleration. It has been studied in detail by Born, already in 1909, who called it ``hyperbolic motion" since the equation of the trajectory in the $x,t$ plane (spacetime) is a hyperbola \cite{born1909theory, barrett2019hyperbolic}.\\

The basic physical concept in a hyperbolic frame is the hyperbolic velocity, also known as the rapidity, given by
\begin{equation}
\label{hypvel}
V={\rm arctanh}(v)~,
\end{equation}
where we have considered the speed of light $c=1$ and where $v=r'$, namely, the derivative with respect to laboratory time. \\
The hyperbolic acceleration, $\alpha$, is defined as the derivative of the hyperbolic velocity $V$ with respect to the proper time $s$. Note that
\begin{equation}
\begin{split}
\alpha=\frac{dV}{ds}=\frac{d}{ds}{\rm arctanh}(v)=\frac{dt}{ds}\frac{d}{dt}{\rm arctanh}(v)=\gamma\frac{d}{dt}{\rm arctanh}(v)\\
=\frac{1}{\sqrt{1-v^2}}\frac{1}{1-v^2}\frac{dv}{dt}=\gamma^3(v)\frac{dv}{dt}~,
\end{split}
\end{equation}
so the hyperbolic acceleration, $\alpha$, is the proper acceleration, related to the laboratory acceleration by $\alpha=\gamma^3(v)(dv/dt)$. The relativistic trajectory can be found from
\begin{equation}
\frac{dv}{(1-v^2)^{3/2}}=\alpha dt~.
\end{equation}
With the initial condition $v=0$ at $t=0$, one obtains for $v$
\begin{equation}
v=\frac{\alpha t}{\sqrt{1+\alpha^2t^2}}~,
\end{equation}
and considering that $v=r'=x'=dx/dt$, by one more integration one can obtain the $x$ coordinate
\begin{equation}
x-x_0=\frac{1}{\alpha}(\sqrt{1+(\alpha t)^2}-1)~.
\end{equation}
The latter equation can be written as the equation of a hyperbola
\begin{equation}
(x-x_0+b)^2-t^2=b^2~,
\end{equation}
where $b=1/\alpha$. We set the initial condition to $x_0=0$ and use the hyperbolic parametrization
\begin{equation}
\label{parsolpart}
x=b(\cosh u-1)~,\ \ \ t=b\sinh u~,
\end{equation}
so that
\begin{equation}
dx=b\sinh u du~,\ \ \ dt=b\cosh udu~,
\end{equation}
which allows to identify $u$ with the rapidity $V$, because
\begin{equation}
\frac{dx}{dt}=\tanh u~.
\end{equation}
Furthermore
\begin{equation}
ds=\sqrt{dt^2-dx^2}=bdu=bdV~,
\end{equation}
so that
\begin{equation}
\label{ec13}
\frac{dV}{ds}=\frac{du}{ds}=\frac{1}{b}=\alpha~.
\end{equation}
Equation (\ref{parsolpart}) contains the components of the position $1+1$-vector
\begin{align}
\label{xsolpart}
X(s)=\frac{1}{\alpha}\begin{pmatrix}
          \sinh(u(s)) \\
          \cosh(u(s)) -1
     \end{pmatrix},
\end{align}
but to generate the iterative sequence of the higher order kinematical quantities, we find that the shifted $1+1$ position vector $\bar{X}$
\begin{equation}
\label{xbarsolpart}
\bar{X}=X+\frac{1}{\alpha}\begin{pmatrix}
          0 \\
          1
     \end{pmatrix}=\frac{1}{\alpha}\begin{pmatrix}
          \sinh(u(s)) \\
          \cosh(u(s))
     \end{pmatrix},
\end{equation}
is more appropiate. In terms of the shifted $1+1$ position vector, the $1+1$-velocity is
\begin{align}
\label{usolpart}
U(s)=\frac{d\bar{X}}{ds}=\begin{pmatrix}
          \cosh u \\
          \sinh u
     \end{pmatrix},
\end{align}
where $\dot{u}=\alpha$ from (\ref{ec13}) has been used. For the $1+1$-acceleration, one obtains
\begin{align}
\label{asolpart}
A(s)=\frac{dU}{ds}=\alpha\begin{pmatrix}
          \sinh u \\
          \cosh u
     \end{pmatrix}.
\end{align}
Furthermore, the $1+1$-Jerk and the $1+1$-Snap are
\begin{align}
\label{jssolpart}
J(s)=\frac{dA}{ds}=\alpha^2\begin{pmatrix}
          \cosh u \\
          \sinh u
     \end{pmatrix},\qquad
S(s)=\frac{dJ}{ds}=\alpha^3\begin{pmatrix}
          \sinh u \\
          \cosh u
     \end{pmatrix}.
\end{align}
The higher order derivatives can be written iteratively as follows:
\begin{equation}
\frac{d^{2p}}{ds^{2p}}\bar{X}(s)=\alpha^{2p}\bar{X},\qquad p=0,1,2,3...
\end{equation}
for even derivatives, and
\begin{equation}
\frac{d^{2q+1}}{ds^{2q+1}}\bar{X}(s)=\alpha^{2q}\frac{d\bar{X}}{ds},\qquad q=0,1,2,3...
\end{equation}
for odd derivatives. For constant proper acceleration, the higher order even derivatives ($1+1$-acceleration, $1+1$-Snap, etc.) are given in terms of rescaled $1+1$-position vectors while the odd derivatives are in terms of rescaled $1+1$-velocities. The moduli of both types of derivatives are succesive powers of fourth order in $\alpha$. The even derivatives are spacelike $1+1$-vectors, whereas the odd ones are timelike. 
There is not much functional change when passing from one kinematic derivative to the next one, which consists only in a permutation of the hyperbolic functions from the time component to the space component at each order and a power-law modification of the square modulus.

\section{Hyperbolic motion in the $1+3$ dimensional Minkowski spacetime}

When considering relativistic hyperbolic motion in more dimensions it is suitable to use the matrix formalism since Lorentz transformations, such as general Lorentz boosts, have well defined matrix representations. In the case of proper acceleration and proper velocity, one can use the following matrix relationship \cite{pons2019observers}
\begin{align}
\label{ec1}
A_{pr}(s)=-\dot{B}(s)B^{-1}(s)U_{pr}=\begin{pmatrix}
           0 \\
           a_x \\
           a_y \\
           a_z
         \end{pmatrix},
\end{align}
where the laboratory velocity is defined as
\begin{align}
U(s)=\gamma\begin{pmatrix}
           1 \\
           \vec{r}\,'(t(s))
         \end{pmatrix},
\end{align}
and is related to proper velocity by the boost transformation
\begin{align}
U(s)\to B(s)U(s)=\begin{pmatrix}
           1 \\
           0 \\
           0 \\
           0
         \end{pmatrix}=U_p,
\end{align}
(the traveler is at rest in the proper frame) and $B(s)$ is a matrix of the family of Lorentz transformations
\begin{align}
\label{BLorentza}
B(s)=\begin{pmatrix}
           \gamma & -\gamma x'(t) & -\gamma y'(t) & -\gamma z'(t) \\
           -\gamma x'(t) & 1+\frac{\gamma^2}{\gamma+1}x'(t)^2 & \frac{\gamma^2}{\gamma+1}x'(t)y'(t) & \frac{\gamma^2}{\gamma+1}x'(t)z'(t) \\
	   -\gamma y'(t) & \frac{\gamma^2}{\gamma+1}x'(t)y'(t) & 1+\frac{\gamma^2}{\gamma+1}y'(t)^2 & \frac{\gamma^2}{\gamma+1}y'(t)z'(t) \\
           -\gamma z'(t) & \frac{\gamma^2}{\gamma+1}x'(t)z'(t) & \frac{\gamma^2}{\gamma+1}y'(t)z'(t) & 1+\frac{\gamma^2}{\gamma+1}z'(t)^2
         \end{pmatrix},
\end{align}
where $t=t(s)$ and the prime indicates derivative with respect to the laboratory time, $t$. From (\ref{ec1}) we obtain
\begin{equation}
\label{vecacel}
\vec{a}=\gamma\vec{r}\,''+\frac{\gamma}{\gamma+1}\gamma'\vec{r}\,',
\end{equation}
where $\gamma'=(1/2)\gamma^3d(\vec{r}\,')^2/dt$ and $\gamma=dt/ds$. Here, $\vec{r}$ is the radius vector and the point indicates derivative with respect to proper time. One can notice that if one takes $\gamma'=0$ in (\ref{vecacel}) then one has $d(\vec{r}\,')^2/dt=0$ and then (\ref{vecacel}) turns into $\vec{a}=0$, which tells us that the motion is a simple translation of constant velocity. Also, we get that the square modulus of the four-acceleration is
\begin{equation}
A^2=A_{pr}^2=-\vec{a}\cdot\vec{a}=-(\gamma^4(\vec{r}\,'')^2+(\gamma')^2),
\end{equation}
where the minus sign comes from the chosen signature ($+---$). We study the case with constant proper acceleration, $\vec{a}$, of constant square modulus $\vec{a}\,^2=\alpha^2$. Considering $\gamma=\dot{t}$, equation (\ref{vecacel}) becomes
\begin{equation}
\ddot{\vec{r}}(s)=\vec{a}+\frac{\ddot{t}(s)}{\dot{t}(s)+1}\dot{\vec{r}}(s).
\end{equation}
We parameterize the four-velocity as
\begin{align}
\label{vel}
U(s)=\begin{pmatrix}
           \dot{t}(s) \\
           \dot{\vec{r}}(s)
         \end{pmatrix}=
\begin{pmatrix}
           \cosh(f(s)) \\
           \sinh(f(s))\hat{n}(s)
         \end{pmatrix},
\end{align}
where $\hat{n}$ is a unitary vector, whose derivative is
\begin{equation}
\label{ec17}
\dot{\hat{n}}=\frac{1}{\sinh(f)}(\vec{a}-\dot{f}\hat{n}).
\end{equation}
Because $\dot{\hat{n}}\cdot\hat{n}=0$, we obtain
\begin{equation}
\label{ec18}
\dot{f}=\vec{a}\cdot \hat{n} \ \ {\rm and} \ \ \ddot{f}=\vec{a}\cdot\dot{\hat{n}},
\end{equation}
which leads to the second order nonlinear equation
\begin{equation}
\label{ec19}
\sinh(f)\ddot{f}+\dot{f}^2-\alpha^2=0.
\end{equation}
A particular solution of this equation corresponding to the initial velocity $U^T(0)=(1,0)$ is
\begin{equation}
\label{ec20}
f_p(s)=\alpha s,
\end{equation}
with $\dot{\hat{n}}_p=0$, $\vec{a}\cdot\vec{a}=\alpha^2$ and for this particular case, $\hat{n}_p=\vec{a}/\alpha$. However, we move to calculate the $1+3$-acceleration, Jerk, and Snap, mantaining an unspecified functional form of $f(s)$ only constrained by equation (\ref{ec19}) to allow for the possibility of using the general solution instead of the particular one (\ref{ec20}).

\subsection{The four-Acceleration}

In the multidimensional space $\hat{n}$ is not necessarily proportional to $\vec{a}$, and in general is a non-constant vector related to $\dot{f}$ by equations (\ref{ec18}), which are important for later calculations. The particular solution $f_p(s)=\alpha s$ is equivalent to $\vec{a}-\hat{n}_p\dot{f}_p=0$, while in general $\vec{a}-\hat{n}\dot{f}\neq 0$. For a two-dimensional Minkowski system we can omit $\hat{n}$ as in the start of the previous section, but for more dimensions we keep a general $\hat{n}$.

We already have the velocity in terms of the general solution $f(s)$, (\ref{vel}), and so the calculations for the acceleration are straightforward with $A=dU/ds$
\begin{align}
\label{ec217}
A(s)=
\begin{pmatrix}
           \dot{f}\sinh (f) \\
           \dot{f}\cosh (f)\hat{n}(s)+\sinh(f)\dot{\hat{n}}
         \end{pmatrix}
=\dot{f}\begin{pmatrix}
           \sinh (f) \\
           \cosh(f)\hat{n}
 \end{pmatrix}
+
\begin{pmatrix}
           0 \\
           \vec{a}-\dot{f}\hat{n}
\end{pmatrix},
\end{align}
where in the last step (\ref{ec17}) has been used. 
Contrary to $\vec{a}$, although $\hat{n}$ is a unitary vector, it is not constant with respect to proper time, as it may change according to (\ref{ec17}).\\
Differentiating $U^2=1$ with respect to proper time we obtain $U\cdot A=0$, which shows the basic fact that relativistic velocity and acceleration are always orthogonal to each other.

Now, let us compute the square modulus of the four-acceleration that we expect to be $-\alpha^2$ due to proper acceleration only having spatial components. 
As $A^2=A_t^2-\vec{A}\,^2$, we obtain
\begin{equation}
A^2=\dot{f}^2\sinh^2 f-\alpha^2-\dot{f}^2\cosh^2f+\dot{f}^2=-\alpha^2.
\end{equation}

\subsection{The four-Jerk}

Jerks are expected to be of no importance in the dynamics of particles in classical mechanics since Newton's equation of motion is of the second order $F(x,\dot{x})=m\ddot{x}$. However, the force depends on $x$ or $\dot{x}$ at previous times, in other words, has memory, then, if we take the time derivative of the equation of motion we have a Jerk dynamics that will contain extra terms. Such models are used in chaos theories. More generally, in classical physics,
the Jerk dynamics have been studied in specific two- and three-dimensional cases such as \cite{flores2020planar,hu2008perturbation,eichhorn1998transformations}, and its applications have been investigated in crystals with memory \cite{te2021jerky}, geomagnetic Jerks \cite{bloxham2002origin} and other chaotic systems \cite{sprott1997some}. We also mention the famous example of a third-order equation of motion in classical electrodynamics as embodied by the Lorentz-Dirac equation of the relativistic electron \cite{poisson1999introLD}, which is closer to the hyperbolic Jerk that we discuss here.\\

In the four-dimensional Minkowski spacetime, we compute the four-jerk as the derivative of the four-acceleration (\ref{ec217}) with respect to proper time
\begin{align}
\label{ec41}
J(s)=\frac{dA}{ds}=
\begin{pmatrix}
           \ddot{f}\sinh (f) + \dot{f}^2\cosh(f) \\
           \dot{f}(\cosh(f)-1)\dot{\hat{n}}+[\ddot{f}(\cosh(f)-1)+\dot{f}^2\sinh(f)]\hat{n}
         \end{pmatrix}.
\end{align}
Considering equations (\ref{ec17}) and (\ref{ec18}), the expression for the Jerk is
\begin{align}
\label{ecJerk}
J(s)=
\begin{pmatrix}
           \alpha^2+\dot{f}^2(\cosh f-1) \\
           \frac{\cosh f-1}{\sinh f}\left\{[\alpha^2+\dot{f}^2(\cosh f-1)]\hat{n}+\dot{f}\vec{a}\right\}
         \end{pmatrix},
\end{align}
which can also be written as
\begin{align}
\label{ecJerkc}
J(s)=\dot{f}^2
\begin{pmatrix}
           \cosh(f) \\
           \sinh(f) \hat{n}
         \end{pmatrix}+
\begin{pmatrix}
           \alpha^2-\dot{f}^2 \\
           \frac{\cosh f -1}{\sinh f}\left[(\alpha^2-2\dot{f}^2)\hat{n}+\dot{f}\vec{a}\right]
         \end{pmatrix}.
\end{align}
The total square modulus, $J^2=J_t^2-\vec{J}\,^2$, is
\begin{equation}
\label{ec45}
J^2=\alpha^4\frac{2+(\dot{f}/\alpha)^2(\cosh f -1)}{\cosh f +1},
\end{equation}
where $\hat{n}\cdot\vec{a}=\dot{f}$ has been used. Derivating twice $U^2=1$ we get $U\cdot J+A^2=0$ so that $U\cdot J=-A^2=\alpha^2$, and we can check that it is satisfied for (\ref{ec217}) and (\ref{ec41}). 
We also check that the total modulus of the Jerk is constant by calculating the derivative of (\ref{ec45}) and proving it to be zero. We find that
\begin{equation}
\label{ec46}
\frac{d}{ds}(J^2)=\frac{2\alpha^2\dot{f}\sinh(f)}{(\cosh f+1)^2}(\ddot{f}\sinh(f)-\alpha^2+\dot{f}^2)~,
\end{equation}
where the last parenthesis in the right hand side is zero by (\ref{ec19}).

\subsection{The four-Snap}
In classical physics, even much rarer systems are the systems where the Snap is taken into account, because being the derivative of the Jerk, it can show up in an even more transient and localized way than the Jerk. In fact, we have been able to find only one paper in which the authors have demonstrated that Snap could be of importance in some chaotic electric systems \cite{leutcho2018unique}.\\

In Minkowski spacetime, the relativistic Snap is defined as the proper time derivative of the Jerk
\begin{align}
\label{ec27}
S(s)=
\frac{d}{ds}\begin{pmatrix}
          \alpha^2+\dot{f}^2(\cosh f-1) \\
           \frac{\cosh f-1}{\sinh f}\left\{[\alpha^2+\dot{f}^2(\cosh f-1)]\hat{n}+\dot{f}\vec{a}\right\}
         \end{pmatrix},
\end{align}
which, considering (\ref{ec17}) and (\ref{ec19}), can be written as
\begin{align}
\label{ec30}
S(s)=
\begin{pmatrix}
           \tanh\left(\frac{f}{2}\right)\dot{f}(2\alpha^2+\dot{f}^2(\cosh f-1)) \\
           \frac{1}{2}{\rm sech}^2\left(\frac{f}{2}\right)(2\alpha^2+\dot{f}^2(\cosh f-1))\left(\vec{a}+2\dot{f}\sinh^2\left(\frac{f}{2}\right)\hat{n}\right)
         \end{pmatrix}.
\end{align}
We can also write (\ref{ec30}) as
\begin{align}
\label{snapform}
S(s)=\dot{f}^3
\begin{pmatrix}
           \sinh(f) \\
           \cosh(f)\hat{n}
         \end{pmatrix}
+\begin{pmatrix}
           2\dot{f}(\alpha^2-\dot{f}^2)\tanh\left(\frac{f}{2}\right) \\
           \frac{2\alpha^2+\dot{f}^2(\cosh f -1)}{\cosh f +1}\vec{a}+\dot{f}\left(\frac{(2\alpha^2+\dot{f}^2(\cosh f -1))\sinh^2f}{(\cosh f +1)^2}-\dot{f}^2\cosh f\right)\hat{n}
         \end{pmatrix}.
\end{align}
Having the expression for the four-Snap, we can calculate the modulus and its derivative to see if it is constant, just as the modulus of acceleration and Jerk. The square modulus, $S_t^2-\vec{S}\cdot\vec{S}$, is given by
\begin{equation}
\label{s2gen}
S^2=-\frac{1}{4}{\rm sech}^4\left(\frac{f}{2}\right)\left(2\alpha^3+\alpha\dot{f}^2(\cosh f-1)\right)^2~,
\end{equation}
where we used $\hat{n}\cdot\hat{n}=1$ and $\vec{a}\cdot\hat{n}=\dot{f}$. One can easily check that $dS^2/ds=0$.
One can go on with calculating the derivatives of each modulus of the subsequent derivatives (Crackle, Pop, etc.), but it is clear that the expected result is that they all vanish. We also notice that if the particular solution $f(s)=f_p(s)=\alpha s$ is used then (\ref{ec217}), (\ref{ecJerkc}) and (\ref{snapform}) turn into (\ref{asolpart}) and (\ref{jssolpart}), respectively.

\section{The general solution to the nonlinear equation for $f(s)$} 

Equation (\ref{ec19}) is vital to obtain the expressions for acceleration, Jerk, and Snap in their simplified form containing only terms of $f$ and $\dot{f}$. Besides, the derivatives of their corresponding moduli vanish only when we consider (\ref{ec19}). To find a solution to (\ref{ec19}), we note that this is a non-homogeneous differential equation of second order where the $\sinh(f(s))$ factor complicates the equation. This can be avoided by the following change of dependent variable
\begin{equation}
\label{flogg}
f(s)=\ln(g(s)),
\end{equation}
which turns (\ref{ec19}) into 
\begin{equation}
\label{ecdiffln}
g(g^2-1)\ddot{g}-(g^2-2g-1)\dot{g}^2-2\alpha^2g^3=0.
\end{equation}
With a second change of the dependent variable
\begin{equation}
\label{ecgarcoth}
g(s)=\frac{1+h(s)}{1-h(s)},
\end{equation}
in (\ref{ecdiffln}), one obtains
\begin{equation}
4h(h^2-1)\ddot{h}-4(h^2+1)\dot{h}^2-\alpha^2(h^2-1)^3=0.
\end{equation}
The general solution of the latter equation is
\begin{equation}
\label{echartanh}
h_\pm(s)=\pm\frac{1}{\alpha}\sqrt{\beta^2+(\alpha^2-\beta^2)\tanh^2\left(\frac{1}{2}\sqrt{\alpha^2-\beta^2}(s-s_0)\right)},
\end{equation}
where $s_0$ and $\beta$ are integration constants that can be determined through initial conditions. Futhermore, we will use the shorthand notations $s_{\alpha\beta}=\frac{1}{2}\sqrt{\alpha^2-\beta^2}(s-s_0)$ and ${\rm Th}^2(s_{\alpha\beta})=(\alpha^2-\beta^2)\tanh^2(s_{\alpha\beta})$; replacing (\ref{echartanh}) in (\ref{ecgarcoth}) gives $g(s)$ in the form
\begin{equation}
g(s)_\pm=\frac{\alpha\pm\sqrt{\beta^2+{\rm Th}^2(s_{\alpha\beta})}}{\alpha\mp\sqrt{\beta^2+{\rm Th}^2(s_{\alpha\beta})}},
\end{equation}
which implies
\begin{equation}
\label{gensolc}
f(s)_\pm=\ln\left(\frac{\alpha\pm\sqrt{\beta^2+{\rm Th}^2(s_{\alpha\beta})}}{\alpha\mp\sqrt{\beta^2+{\rm Th}^2(s_{\alpha\beta})}}\right),
\end{equation}
or alternatively
\begin{equation}
\label{ec33}
f(s)_\pm=\pm 2{\rm arctanh}\left(\frac{1}{\alpha}\sqrt{\beta^2+{\rm Th}^2(s_{\alpha\beta})}\right).
\end{equation}
If $\beta=0$, then the general solution (\ref{ec33}) reduces to the particular one, $f_p(s)=\alpha (s-s_0)$.\\
Considering (\ref{ec33}) in (\ref{ec45}), the square modulus of the Jerk is obtained in terms of $\alpha$ and the initial condition $\beta$
\begin{equation}
\label{modjerkab}
J^2=\alpha^2(\alpha^2-\beta^2),
\end{equation}
which shows that the square modulus of the Jerk is a constant quantity less than $\alpha^4$. Regarding the square modulus of the four-Snap in (\ref{s2gen}), one obtains
\begin{equation}
\label{snapmodulus}
S^2=-\alpha^2 (\alpha^2 -\beta^2 )^2=-\frac{J^4}{\alpha^2},
\end{equation}
by using the general solution (\ref{ec33}), which shows that the four-Snap is spacelike of square modulus less than $\alpha^6$. One can also write the general solution (\ref{ec33}) in terms of the Jerk modulus (\ref{modjerkab}) as
\begin{equation}
\label{gensolJ}
f(s)_\pm=\pm 2{\rm arctanh}\left(\frac{1}{\alpha}\sqrt{\left(\alpha^2-\frac{J^2}{\alpha^2}\right)+\frac{J^2}{\alpha^2}\tanh^2\left(\frac{1}{2}\frac{|J|}{\alpha}(s-s_0)\right)}\right).
\end{equation}
Formulas of the general solution in terms of the moduli of higher order derivatives can be easily written down. For instance, using (\ref{snapmodulus}) one can substitute $J^2$ in (\ref{gensolJ}) by the modulus of the Snap.
\\

Due to the performed changes of variable of $f(s)$ and $g(s)$, there are some restrictions we should take into account. Equation (\ref{flogg}) implies that $g(s)\geq 1$, while equation (\ref{ecgarcoth}) requires that $h(s)\in [0,1)$. Since, $0\leq\tanh^2(s_{\alpha\beta})\leq 1$ we make sure that $h(s)\in [0,1)$ only if $|\beta|<|\alpha|$, and the limiting case $h(s)\to 1$ happens when $s\to\infty$ or $|\alpha|=|\beta|$.

\subsection{Kinematical quantities using the general solution $f(s)_{\pm}$} 

Considering equation (\ref{ec33}), the kinematical quantities computed in previous sections can be expressed in terms of the scaled variable $s_{\alpha\beta}$. For the four-acceleration in (\ref{ec217}), one has
\begin{align}
A(s)=
\frac{\alpha{\rm Th}(s_{\alpha\beta})}{\alpha^2-\beta^2}
\begin{pmatrix}
           2\alpha\cosh^2(s_{\alpha\beta})\\
		\\
           \frac{\beta^2+\alpha^2\cosh(2s_{\alpha\beta})}{\sqrt{\beta^2+{\rm Th}^2(s_{\alpha\beta})}}\hat{n}
 \end{pmatrix}
+\alpha
\begin{pmatrix}
           0 \\
		\\
           \hat{a}-\frac{{\rm Th}(s_{\alpha\beta})}{\sqrt{\beta^2+{\rm Th}^2(s_{\alpha\beta})}}\hat{n}
\end{pmatrix}.
\end{align}
For $\beta=0$, the first term reduces to (\ref{asolpart}), while the second term vanishes.\\

For the four-Jerk, one obtains
\begin{equation}
\begin{split}
J(s)=
\frac{\alpha^2{\rm Th}^2(s_{\alpha\beta})}{(\alpha^2-\beta^2)(\beta^2+{\rm Th}^2(s_{\alpha\beta}))}
\begin{pmatrix}
          \beta^2+\alpha^2\cosh(2s_{\alpha\beta})\\
		\\
          2\alpha\cosh^2(s_{\alpha\beta})\sqrt{\beta^2+{\rm Th}^2(s_{\alpha\beta})}\hat{n}
 \end{pmatrix}
+\alpha
\begin{pmatrix}
           \frac{\alpha\beta^2}{\beta^2+{\rm Th}^2(s_{\alpha\beta})} \\
		\\
          {\rm Th}(s_{\alpha\beta})\hat{a}+\frac{\beta^2-{\rm Th}^2(s_{\alpha\beta})}{\sqrt{\beta^2+{\rm Th}^2(s_{\alpha\beta})}}\hat{n}
\end{pmatrix}.
\end{split}
\end{equation}
For $\beta=0$, the first term reduces to the left part of (\ref{jssolpart}), while the second term vanishes.\\

For the Snap
\begin{equation}
\begin{split}
&S(s)=
\frac{\alpha^3{\rm Th}^3(s_{\alpha\beta})}{\alpha^2-\beta^2}
\begin{pmatrix}
          \beta^2+\alpha^2\cosh(2s_{\alpha\beta})\\
		\\
           2\alpha\sqrt{\beta^2+{\rm Th}^2(s_{\alpha\beta})}\cosh^2(s_{\alpha\beta})\hat{n}
 \end{pmatrix}
+\\
&\frac{1}{\cosh^2(s_{\alpha\beta})}
\begin{pmatrix}
           0\\
		\\
           \left[(\alpha^2-\beta^2)+\frac{{\rm Th}^2(s_{\alpha\beta})}{\beta^2+{\rm Th}^2(s_{\alpha\beta})}(\beta^2+\alpha^2\sinh^2(s_{\alpha\beta}))\right]\vec{a}
\end{pmatrix}
+\\
&\frac{\alpha{\rm Th}(s_{\alpha\beta})}{\sqrt{\beta^2+{\rm Th}^2(s_{\alpha\beta})}}
\begin{pmatrix}
          \frac{2\alpha\beta^2}{\sqrt{\alpha^2-\beta^2}}\frac{1}{\sqrt{\beta^2+{\rm Th}^2(s_{\alpha\beta})}}  \\
		\\
          \left[\frac{\alpha^2-\beta^2}{\cosh s_{\alpha\beta}}+\frac{(\alpha^2-\beta^2){\rm Th}^4(s_{\alpha\beta})}{2\cosh^4 s_{\alpha\beta}}\frac{\beta^2+\alpha^2\sinh^2s_{\alpha\beta}}{(\beta^2+{\rm Th}^2(s_{\alpha\beta}))^2}-\frac{\alpha^2{\rm Th}^2(s_{\alpha\beta})}{\beta^2+{\rm Th}^2(s_{\alpha\beta})}\frac{\beta^2+\alpha^2\cosh(2s_{\alpha\beta})}{\alpha^2-\beta^2}\right]\hat{n}
\end{pmatrix}.
\end{split}
\end{equation}
For $\beta=0$, the first term reduces to the right part of (\ref{jssolpart}), while the second and third term eliminate each other.

\section{Modified Rindler hyperbolas}

\begin{figure}[h]
\centering
\includegraphics[scale=0.7]{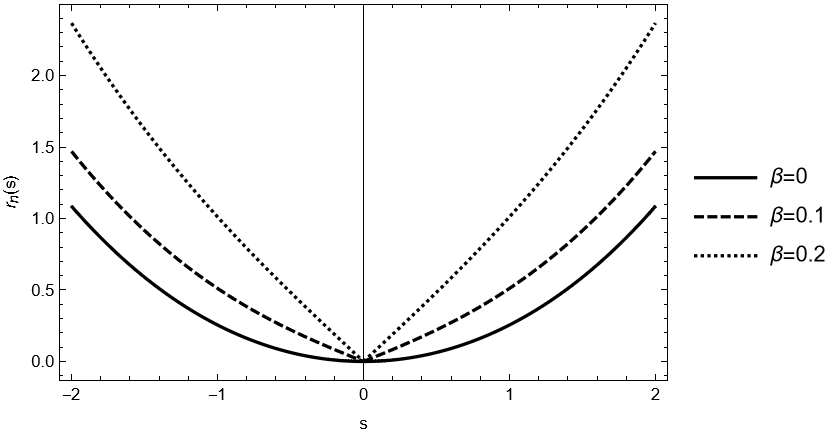}
\caption{Plots of $r_n$ vs $s$, with $\alpha=0.5$, for the standard Rindler hyperbola ($\beta=0$) and the deformed `hyperbolas'
corresponding to $\beta=0.1$ and $\beta=0.2$.}
\label{rsbeta}
\end{figure}

\begin{figure}[h]
\centering
\includegraphics[scale=0.7]{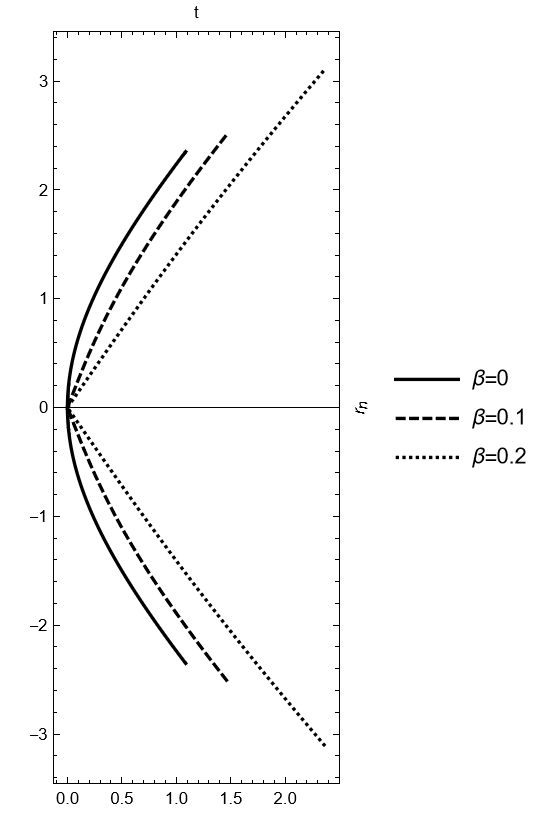}
\caption{For fixed $\alpha=0.5$, original Rindler hyperbola ($\beta=0$) and modified hyperbolas for $\beta=0.1$, and $\beta=0.2$, respectively.}
\label{pprsbeta}
\end{figure}

If one uses the particular solution $f_p(s)$, it is easy to obtain the standard geometric hyperbolic behavior in the plane defined by the coordinates $t(s)$ and $x(s)$ by integrating the corresponding components of the parametrized velocity (\ref{vel}). Instead of this, we use the general arctanh solution to see graphically the kind of geometric behavior of the worldlines. Using (\ref{vel}) and (\ref{ec33}) we compute the coordinate time as
\begin{equation}
t(s)=\int \cosh(f(s))ds=\left(1-\frac{\beta^2}{\alpha^2}\right)^{-3/2}\frac{\sinh(2s_{\alpha\beta})}{\alpha}+\frac{\frac{\beta^2}{\alpha^2}}{1-\frac{\beta^2}{\alpha^2}}s.
\end{equation}
Similarly, by integrating the space component of the velocity  along $\hat{n}$, which is $\sinh(f)$, leads to the projection $r_n$
\begin{equation}
\begin{split}
r_n&(s)=\int \sinh(f(s))ds\\
&=\frac{2}{(\alpha^2-\beta^2)^{3/2}}\left(\alpha\sinh(s_{\alpha\beta})\sqrt{\beta^2+\alpha^2\sinh^2(s_{\alpha\beta})}+\beta{\rm arcsinh}\left(\frac{\alpha\sinh(s_{\alpha\beta})}{\beta}\right)\right).
\end{split}
\end{equation}
The plot of $r_n(s)$ as a function of $s$ is displayed in fig. (\ref{rsbeta}), for different values of $\beta$, whereas in the $t$ vs $r_n$ coordinates as frequently presented in the literature \cite{born1909theory,rindler1966kruskal} are displayed in fig. (\ref{pprsbeta}). The value $\beta=0$ corresponds to the particular solution $f_p(s)=\alpha s$.

We notice that one can also work with the hyperbolic tangent of the equivalent of the hyperbolic velocity (rapidity) as defined in (\ref{hypvel}), where we recall that the four-velocity can be represented as $U^T=\gamma(1,\vec{r}\,')$, and we have for the hyperbolic parametrization that
\begin{equation}
U^T(s)=(\cosh(f(s)),\sinh(f(s))\hat{n}),
\end{equation}
then
\begin{equation}
\gamma=\frac{dt}{ds}=\frac{1}{\sqrt{1-(r')^2}}=\cosh(f(s)),
\end{equation}
where we can obtain
\begin{equation}
\label{threevelmag}
|r'|=\tanh(f(s)),
\end{equation}
the modulus of the three-velocity. Considering equation (\ref{ec33}), and considering that
\begin{equation}
\tanh(2 {\rm arctanh}(x))=\tanh\left(\ln\left(\frac{1+x}{1-x}\right)\right)=\frac{2x}{1+x^2},
\end{equation}
we can write (\ref{threevelmag}) as
\begin{equation}
|r'|=\tanh(f(s))=\alpha\frac{1+\cosh(2s_{\alpha\beta})}
{\beta^2+\alpha^2\cosh(2s_{\alpha\beta})}\sqrt{\beta^2+{\rm Th}^2(s_{\alpha\beta})}~.
\end{equation}
The results are equivalent as $|r'|=\tanh(f)=\dot{r}/\gamma=\dot{r}/\cosh(f)$, so $\dot{r}(s)=\sinh(f)$.
However, it is of physical interest because $|r'|$ is the  magnitude of the three-velocity, the velocity that the laboratory frame measures for the moving observer.

\section{Conclusions}

We have studied the kinematical quantities of relativistic hyperbolic motion, i.e., of constant proper acceleration $\alpha$, in $1+1$- and $1+3$-dimensional Minkowski spacetime. The standard hyperbolic parametrization of the spacetime coordinates has been used in the literature to obtain
a nonlinear differential equation for the argument of the hyperbolic functions.
 In this paper, we have worked both with the particular linear solution as in the recent literature \cite{pons2019observers}, but also with the general ${\rm arctanh}$ solution to evaluate the kinematical higher quantities. All of the higher order derivatives beyond the acceleration depend only on the proper constant acceleration when the particular solution is employed, but if the general solution is used, they depend also on an integration constant corresponding to a nonzero initial condition. In the physics context, the effect of this nonzero initial condition is to produce deformed Rindler hyperbolas which still belong to the class of relativistic hyperbolic motion since all the proper time derivatives are of constant square modulus, although of smaller weight than in the case of the particular solution.

\bibliographystyle{ieeetr}
\bibliography{references} 

\begin{thebibliography}{n}

\bibitem{unruh1976notes}
W.~G.~Unruh,
Notes on black hole evaporation,
{\em Phys. Rev. D} \textbf{14} (1976) 870-892. https://doi:10.1103/PhysRevD.14.870

\bibitem{crispino2008unruh}
L.~C.~B.~Crispino, A.~Higuchi and G.~E.~A.~Matsas,
The Unruh effect and its applications,
{\em Rev. Mod. Phys.} \textbf{80} (2008) 787-838. https://doi:10.1103/RevModPhys.80.787

\bibitem{russo2009relativistic}
J.~G.~Russo and P.~K.~Townsend,
Relativistic kinematics and stationary motions,
{\em J. Phys. A} \textbf{42} (2009) 445402. https://doi.org/10.1088/1751-8113/42/44/445402

\bibitem{letaw1981stationary}
J.~R.~Letaw,
Vacuum excitation of noninertial detectors on stationary world lines,
{\em Phys. Rev. D} \textbf{23} (1981) 1709-1714. https://doi:10.1103/PhysRevD.23.1709

\bibitem{friedman2015uniform}
Y.~Friedman and T.~Scarr,
Uniform acceleration in general relativity,
{\em Gen. Rel. Grav.} \textbf{47} (2015) 121. https://doi:10.1007/s10714-015-1966-5

\bibitem{paithankar2019bound}
K.~Paithankar and S.~Kolekar,
Bound on Rindler trajectories in a black hole spacetime,
{\em Phys. Rev. D} \textbf{99} (2019) 064012. https://doi.org/10.1103/PhysRevD.99.064012

\bibitem{born1909theory} M. Born, The theory of the rigid electron in the kinematics of the principle of relativity,
{\em Ann. Phys.(Leipzig)} \textbf{335} (1909) 1-56. https://doi.org/10.1002/andp.19093351102

\bibitem{pons2019observers}
J.~M.~Pons and F.~de Palol, Observers with constant proper acceleration, constant proper jerk, and beyond,
{\em Gen. Rel. Grav.} \textbf{51} (2019) 80. https://doi.org/10.1007/s10714-019-2562-x

\bibitem{barrett2019hyperbolic} J. F. Barrett, The hyperbolic theory of special relativity,
arXiv:1102.0462v2 (2019) 109 pages. https://doi.org/10.48550/arXiv.1102.0462

\bibitem{flores2020planar} E. Flores-Gardu\~no, S.C. Mancas, H.C. Rosu, M. P\'erez-Maldonado,
Planar motion with Fresnel integrals as components of the velocity,
{\em Rev. Mex. F\'{\i}s.} \textbf{66} (2020) 585-588. https://doi.org/10.31349/RevMexFis.66.585

\bibitem{hu2008perturbation} H. Hu, Perturbation method for periodic solutions of nonlinear jerk equations,
{\em Phys. Lett. A} \textbf{372} (2008) 1405-1409. https://doi.org/10.1016/j.physleta.2008.03.027.

\bibitem{eichhorn1998transformations} R. Eichhorn, S. J. Linz, and P. H\"anggi, Transformations of nonlinear dynamical systems to jerky
motion and its application to minimal chaotic flows, {\em Phys. Rev. E} \textbf{58} (1998) 7151-7164. https://doi.org/10.1103/PhysRevE.58.7151

\bibitem{te2021jerky} M. te Vrugt, J. Jeggle, and R. Wittkowski, Jerky active matter: a phase field crystal model with
translational and orientational memory, {\em New J. Phys.} \textbf{23} (2021) 063023. https://doi.org/10.1088/1367-2630/abfa61

\bibitem{bloxham2002origin} J. Bloxham, S. Zatman, and M. Dumberry, The origin of geomagnetic jerks, {\em Nature} \textbf{420} 
(2002) 65-68. https://doi.org/10.1038/nature01134

\bibitem{sprott1997some} J. Sprott, Some simple chaotic jerk functions, {\em Am. J. Phys.} \textbf{65} (1997) 537-543. https://doi.org/10.1119/1.18585

\bibitem{poisson1999introLD} E. Poisson, An introduction to the Lorentz-Dirac equation,
arXiv:gr-qc/9912045 (1999) 14 pages. https://doi.org/10.48550/arXiv.gr-qc/9912045

\bibitem{leutcho2018unique} G. D. Leutcho and J. Kengne, A unique chaotic snap system with a smoothly adjustable symmetry
and nonlinearity: Chaos, offset-boosting, antimonotonicity, and coexisting multiple attractors,
{\em Chaos, Solitons \& Fractals} \textbf{113} (2018) 275-293. https://doi.org/10.1016/j.chaos.2018.05.017

\bibitem{rindler1966kruskal} W. Rindler, Kruskal space and the uniformly accelerated frame,
{\em Am. J. Phys.} \textbf{34} (1966) 1174-1178. https://doi.org/10.1119/1.1972547

\end{thebibliography}

\end{document}